\documentclass[a4paper]{jpconf}

\usepackage{aas_macros}
\usepackage[numbers,square,sort&compress]{natbib}
\usepackage{amsmath}
\usepackage{amssymb}
\usepackage{bm}
\usepackage{graphicx}

\begin{document}

\title{The small-scale turbulent dynamo in smoothed particle magnetohydrodynamics}

\author{T S Tricco$^{1,2,3}$, D J Price$^3$ and C Federrath$^{4,3}$}

\address{$^1$ Canadian Institute for Theoretical Astrophysics, University of Toronto, 60 St. George Street, Toronto, ON M5S 3H8, Canada}
\address{$^2$ School of Physics, University of Exeter, Stocker Road, Exeter, EX4 4QL, United Kingdom}
\address{$^3$ Monash Centre for Astrophysics, Monash University, Clayton, VIC 3800, Australia}
\address{$^4$ Research School of Astronomy and Astrophysics, The Australian National University, Canberra, ACT 2611, Australia}

\ead{ttricco@cita.utoronto.ca}

\begin{abstract}
Supersonic turbulence is believed to be at the heart of star formation. We have performed smoothed particle magnetohydrodynamics (SPMHD) simulations of the small-scale dynamo amplification of magnetic fields in supersonic turbulence. The calculations use isothermal gas driven at rms velocity of Mach 10 so that conditions are representative of star-forming molecular clouds in the Milky Way. The growth of magnetic energy is followed for 10 orders in magnitude until it reaches saturation, a few percent of the kinetic energy. The results of our dynamo calculations are compared with results from grid-based methods, finding excellent agreement on their statistics and their qualitative behaviour. The simulations utilise the latest algorithmic developments we have developed, in particular, a new divergence cleaning approach to maintain the solenoidal constraint on the magnetic field and a method to reduce the numerical dissipation of the magnetic shock capturing scheme. We demonstrate that our divergence cleaning method may be used to achieve $\nabla \cdot {\bf B}=0$ to machine precision, albeit at significant computational expense. 
\end{abstract}

\section{Introduction}

There are three broad classes of methods used to simulate astrophysical hydrodynamics. Generally speaking, these are: 
\begin{itemize}
\item Eulerian, grid-based methods which discretise the volume the fluid occupies into a collection of cells. The cells keep their location fixed in space, calculating the transfer of fluid between adjacent cells. Examples of such codes include {\sc Pluto} \cite{pluto1, pluto2}, {\sc Athena} \cite{athena}, {\sc Ramses} \cite{ramses}, and {\sc Flash} \cite{fryxelletal00, dubeyetal08}. 
\item Lagrangian, particle-based methods that discretise the fluid into a set of particles. The particles move according to the fluid flow, mimicing fluid behaviour. The most common approach is smoothed particle hydrodynamics (SPH), with codes such as {\sc Gadget2} \cite{gadget2, ds09}, {\sc Gasoline} \cite{gasoline} and {\sc Phantom} \cite{pf10}. 
\item Methods that combine elements of Eulerian and Lagrangian methods, for example, the moving-mesh code {\sc Arepo} \cite{arepo}, the mesh-free method {\sc Gizmo} \cite{hopkins15}, and the Lagrangian-Eulerian re-map code {\sc Lare3d} \cite{arberetal01}.
\end{itemize}

In this work, we use the second approach, employing smoothed particle magnetohydrodynamics (SPMHD). Right from the beginning when SPH was first formulated in 1977, there was optimism that magnetic fields could be included as part of the method, ``{\it \ldots magnetic fields may be included without difficulty \ldots}" \citep{gm77}. However, this turned out not to be quite so easy in practice -- the conservative SPMHD method is unstable when the magnetic pressure exceeds the thermodynamic pressure \cite{pm85}. This was investigated further over the next 20 years by \citep{morris96, mwd95, bot01} and others, with the modern SPMHD method given in the series of papers by Price \& Monaghan \citep{pm04a, pm04b, pm05}.

Nowadays, SPMHD has achieved a level of maturity that allows for practical simulations, for example, the study of jets \& outflows during protostar formation \cite{ptb12, btp14, lbp15, wpb16}. A significant development has been the constrained hyperbolic divergence cleaning method to handle the divergence-free constraint of the magnetic field \cite{tp12}. 

In this work, we apply SPMHD to the study of small-scale dynamo amplification of magnetic fields in supersonic turbulence, the first time such a regime has been explored with SPMHD. Purely hydrodynamic supersonic turbulence has been studied with SPH by \cite{pf10}, who compared the statistics of driven, supersonic turbulence between SPH and grid-based methods. They found excellent agreement on the qualitative behaviour and statistics of the turbulence. Our current work extends that comparison to include magnetic fields, thus we can be confident that the hydrodynamics are well-modelled and we can focus on differences introduced by magnetic fields.


\section{Recent developments in smoothed particle magnetohydrodynamics}
\label{sec:spmhd}

The set of simulations performed in this work solve the magnetohydrodynamics (MHD) equations, given by
\begin{gather}
\frac{{\rm d}\rho}{{\rm d}t} = - \rho \nabla \cdot {\bf v} , \\
\frac{{\rm d}{\bf v}}{{\rm d}t} = - \frac{1}{\rho} \nabla \left( P + \frac{B^2}{2 \mu_0} \right) + \frac{1}{\mu_0 \rho} \nabla \cdot \left( {\bf B} {\bf B} \right) , \label{eq:momentum} \\
\frac{{\rm d}{\bf B}}{{\rm d}t} = \left( {\bf B} \cdot \nabla \right) {\bf v} - \left( \nabla \cdot {\bf v} \right) {\bf B}, \label{eq:induction} \\
\nabla \cdot {\bf B} = 0 , \label{eq:divbconstraint} 
\end{gather}
which, since we will solve them using a Lagrangian scheme, are written using the Lagrangian time derivative, ${\rm d}/{\rm d}t = \partial / \partial t + ({\bf v} \cdot \nabla)$. The variables have their usual meaning: $\rho$ is the density, ${\bf v}$ is the velocity, ${\bf B}$ is the magnetic field, $P$ is the thermodynamic pressure and $\mu_0$ the permeability of free space. In SPMHD, the density is obtained by summation and solved iteratively with the smoothing length \cite{pm07}. The magnetic tension force is implemented using the approach of \cite{bot01}. A full review of SPMHD is given by \cite{price12, tricco15}.

\subsection{`Constrained' hyperbolic/parabolic divergence cleaning}

The problem with the induction equation (Eq.~\ref{eq:induction}) is that it makes no guarantee to preserve the divergence-free nature of the magnetic field ($\nabla \cdot {\bf B} = 0$). Numerical errors will introduce monopoles into the magnetic field, which may then grow over the long-term and lead to unphysical results. 

We use the `constrained' hyperbolic/parabolic divergence cleaning method \cite{tp12} to remove divergence errors in the magnetic field. The method is an SPMHD adaptation of the method by \cite{dedneretal02}, and cleans the magnetic field by coupling a scalar field, $\psi$, to the magnetic field according to
\begin{gather}
\frac{{\rm d}{\bf B}}{{\rm d}t} = - \nabla \psi , \\
\frac{{\rm d}\psi}{{\rm d}t} = - c_{\rm h}^2 \nabla \cdot {\bf B} - \frac{\psi}{\tau} - \frac{\psi}{2} (\nabla \cdot {\bf v}) . \label{eq:psievo}
\end{gather}
Together, they propagate the divergence of the magnetic field as a damped wave with characteristic speed $c_{\rm h}$ set equal to the local fast MHD wave speed. The damping timescale is $\tau = h / \sigma c_{\rm h}$, where $h$ is the smoothing length and $\sigma=1.0$ in 3D is a dimensionless constant. The third term in Eq.~\ref{eq:psievo} was introduced to account for changes in $\psi$ due to compression and rarefaction of the gas \cite{tp12}.

The method is `constrained' in that it is built using the energy conserving formalism of SPH. Defining the specific energy content of the $\psi$ field to be $e_\psi = \psi^2 / 2 \mu_0 \rho c_{\rm h}^2$, the discretised hyperbolic cleaning equations may be obtained such that the total energy ($e_\psi$ plus magnetic) is constrained to stay constant. This remedies problems that would otherwise occur at sharp density contrasts and free boundaries when the same differential operator is used for $\nabla \psi$ and $\nabla \cdot {\bf B}$ \cite{tp12}, and when coupled with parabolic diffusion, means the divergence cleaning can only ever lead to reductions in divergence error.

\subsection{Aritificial resistivity switch}

Discontinuities in the magnetic field are captured by adding an artificial resistivity \cite{pm04a,pm05}. This smooths the magnetic field on the resolution scale so that it remains single-valued. However, it is equivalent to adding diffusion in the form of $\eta_{\rm AR} \nabla^2 {\bf B}$ with coefficient $\eta_{\rm AR} \sim \alpha_{\rm B} v_{\rm sig} h$, where $\alpha_{\rm B}$ is a dimensionless constant of order unity and $v_{\rm sig}$ is the characteristic speed of the shock. Such diffusion is only meant to treat discontinuities, and the trick is to detect where those discontinuities are located so that artificial resistivity can be switched off when not needed in order to minimise numerical dissipation.

Price \& Monaghan \cite{pm05} used a switch to detect discontinuities, evolving $\alpha_{\rm B}$ using a differential equation with a source term proportional to $\nabla \times {\bf B}$. However, this approach fails for strong shocks with very weak magnetic fields, not setting $\alpha_{\rm B}$ to sufficient levels to keep the magnetic field single-valued \cite{tp13}. 

Our new approach \cite{tp13} is to directly set $\alpha_{\rm B} = h \vert \nabla {\bf B} \vert / \vert {\bf B} \vert$. This measures the relative size of the jump in the magnetic field compared to the absolute value of the field strength, giving a measure which works for any magnetic field strength. This is particularly relevant for the magnetised, supersonic turbulence comparison on which this paper is focused, allowing shocks to be detected during magnetic field amplification \cite{tp13}.

\section{Enhancing the divergence cleaning}
\label{sec:enhanced-cleaning}

The divergence cleaning method can be enhanced in two ways. One is by explicitly increasing the cleaning wave speed. Increasing the wave speed by an over-cleaning factor of 10 would set $c_{\rm h} \rightarrow 10 \times c_{\rm h}$. This approach is easy to implement and has found some use in practice \cite{btp14}, but is computationally expensive because it directly reduces the timestep due to the Courant condition. The second approach is to sub-cycle the cleaning equations between timesteps, pausing the rest of the simulation and holding the particles fixed while the cleaning equations are solved for a number of sub-steps. This is much more computationally efficient than over-cleaning, however is more complicated to implement in code, particularly when dealing with individual timesteps.


\begin{figure}
\centering
\includegraphics[width=0.6\textwidth]{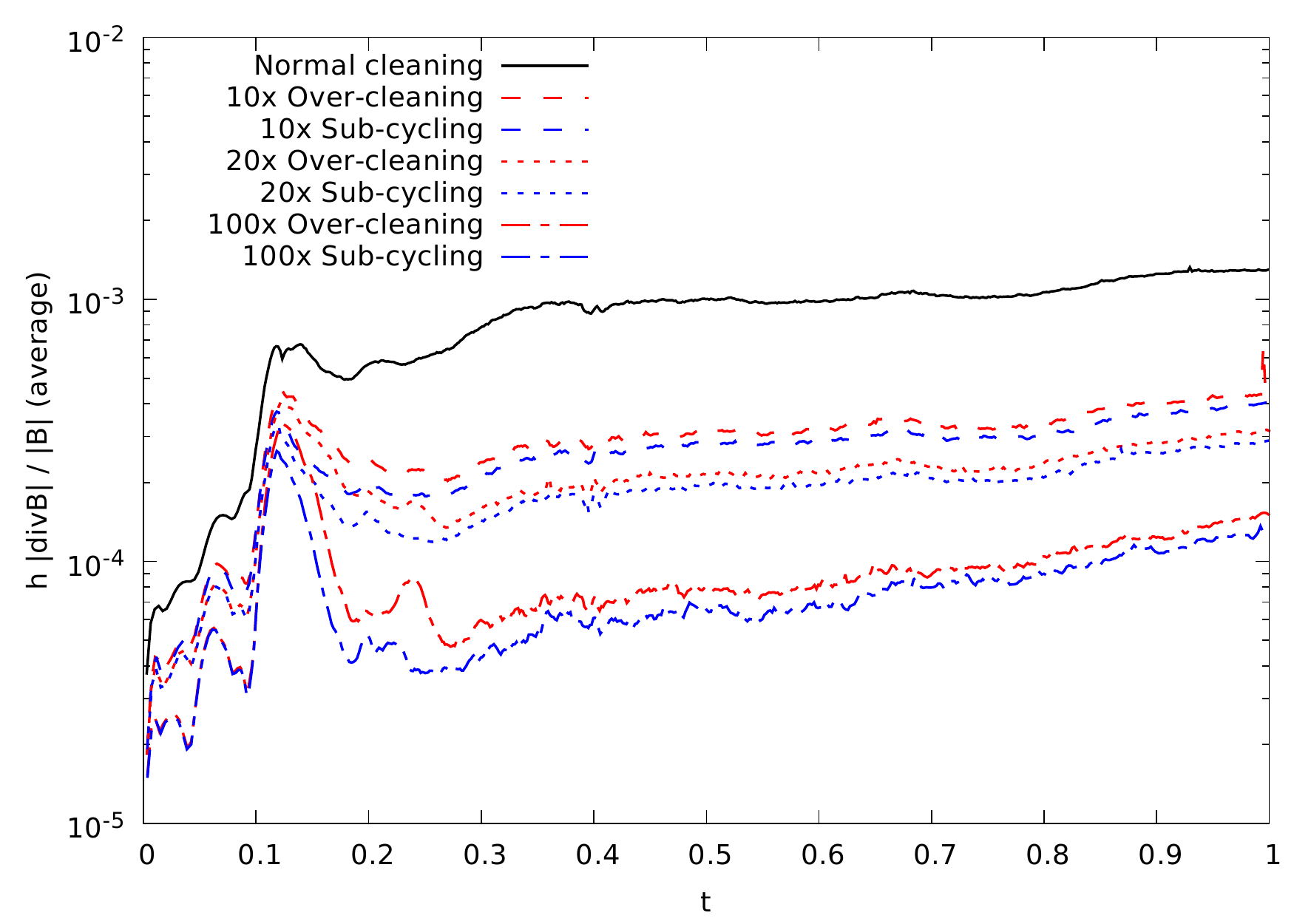}
\caption{Average divergence error for the Orszag-Tang vortex using enhanced divergence cleaning techniques -- 10$\times$, $20\times$, and $100\times$ over-cleaning and sub-cycling with $10$, 20, and 100 iterations. Both approaches yield comparable levels of divergence error when the over-cleaning factor and sub-cycling iteration count are the same.}
\label{fig:ovcitc}
\end{figure}

We have tested over-cleaning and sub-cycling of the divergence cleaning equations using the Orszag-Tang vortex \cite{ot79}. In general, we find that over-cleaning and sub-cycling yield similar results for equivalent over-cleaning factors and numbers of sub-cycles (Fig.~\ref{fig:ovcitc}). That is, an over-cleaning factor of 10 produces results on par with performing 10 sub-cycles every timestep. Each factor of 10 increase in over-cleaning factor and number of sub-cycles produced half an order of magnitude reduction in average $h \vert \nabla \cdot {\bf B} \vert / \vert {\bf B} \vert$ divergence error.

\begin{figure}
\centering
\includegraphics[width=0.45\linewidth]{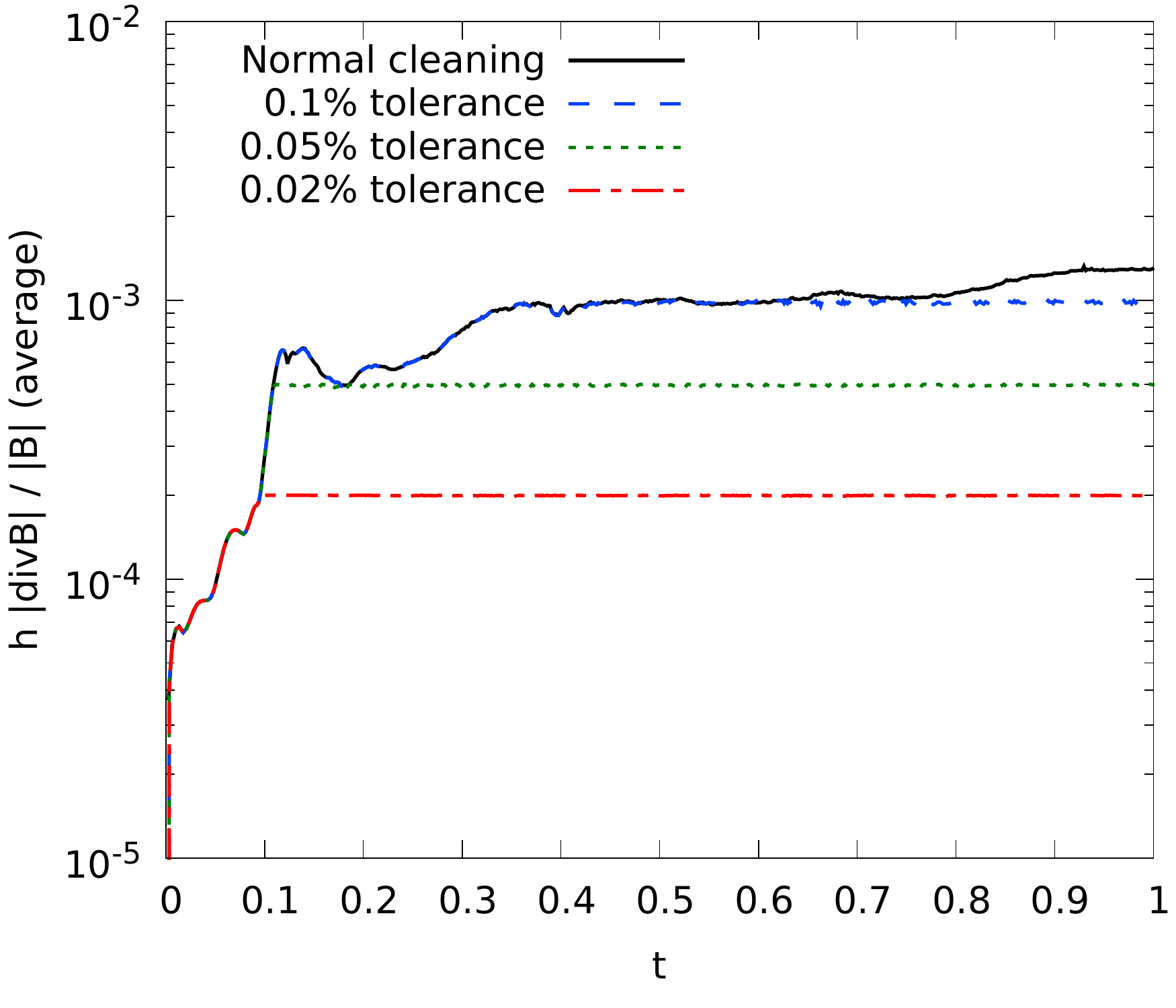}
\includegraphics[width=0.45\linewidth]{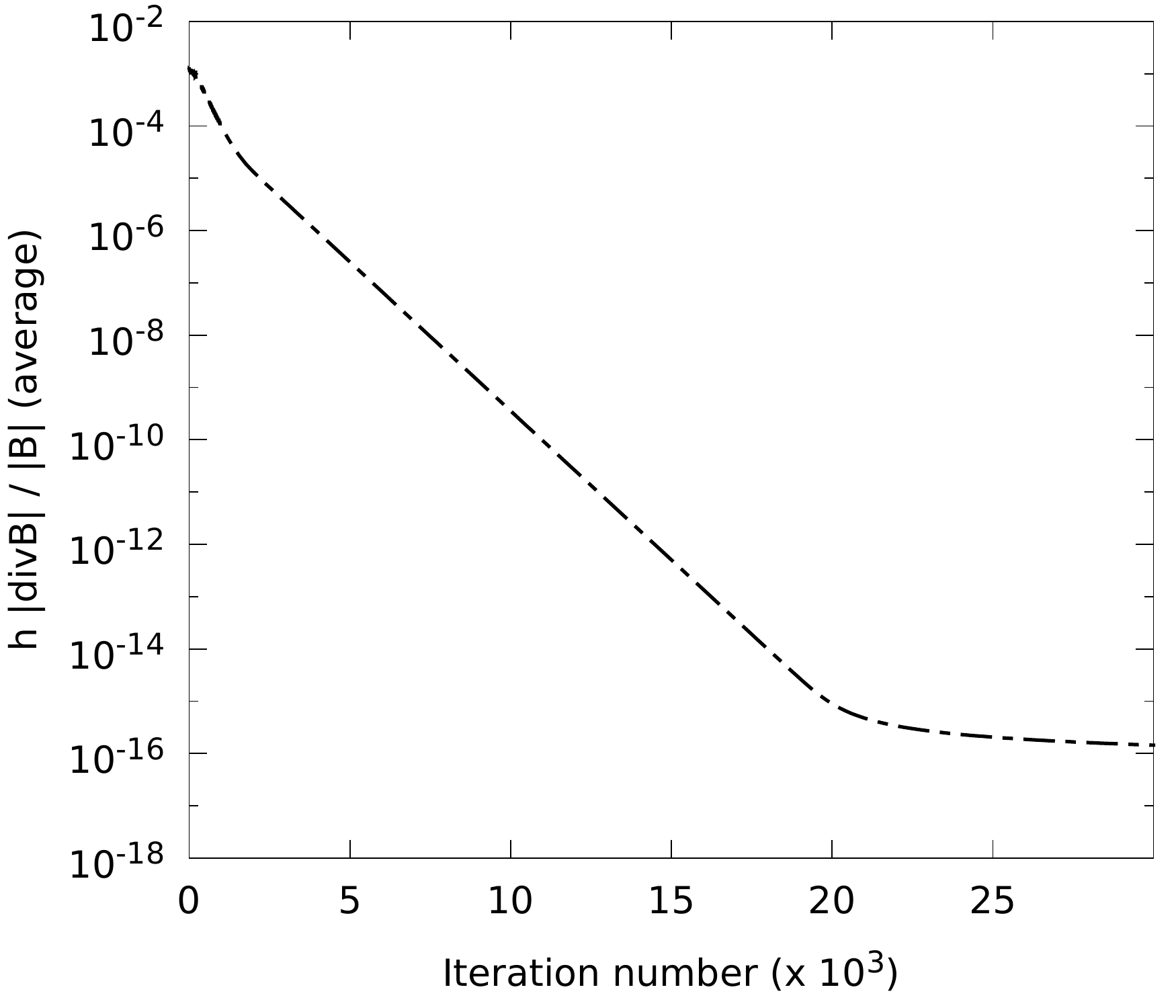}
\caption{{\it Left:} Sub-cycling the divergence cleaning equations is performed each timestep on the Orszag-Tang vortex until the average error is below the specified tolerance, achieving tolerance based divergence cleaning. {\it Right:} The number of iterations required to reduce the average divergence error to specified tolerances. It is possible to clean the divergence of the magnetic field to machine precision ($10^{-16}$--$10^{-18}$), achieving $\nabla \cdot {\bf B}=0$, but to do so requires on the order of a few tens of thousands of iterations.}
\label{fig:itvtol}
\label{fig:itvzero}
\end{figure}

An advantage that sub-cycling offers is tolerance based divergence cleaning. Instead of prescribing a fixed number of iterations, the cleaning equations may be sub-cycled until a particular criterion is met, such as a threshold on average level of divergence error. The left panel of Fig.~\ref{fig:itvtol} shows the average $h \vert \nabla \cdot {\bf B} \vert / \vert {\bf B} \vert$ divergence error with tolerances set to $0.1\%$, $0.05\%$ and $0.02\%$, showing that tolerance based divergence cleaning with sub-cycling is possible. It is also possible to set the tolerance level as strict as desired, and in the right panel of Fig.~\ref{fig:itvzero} we show that `constrained' hyperbolic divergence cleaning may be used to reduce the divergence of the magnetic field all the way to zero, formally achieving $\nabla \cdot {\bf B} = 0$. That is not really feasible in practice though, as the number of sub-cycles required to achieve this is prohibitively expensive on the order of tens of thousands of sub-steps (right panel, Fig.~\ref{fig:itvzero}).

\section{Small-scale turbulent dynamo: SPH compared to Grid}
\label{sec:turb}

Molecular clouds are supersonically turbulent, generating dense filaments that become the birth sites of protostars \cite{mk04, mo07, fk12, padoanetal14}. We have performed simulations of molecular cloud turbulence involving magnetic fields to test our SPMHD methods. The key piece of physics studied is the small-scale dynamo, an example of a {\it fast} dynamo that amplifies magnetic energy at an exponential rate \cite{bs05}.  By stretching, twisting and winding magnetic fields near the dissipation scale (hence `small-scale'), kinetic energy is converted into magnetic energy. 

The growth of the magnetic field due to small-scale dynamo undergoes three phases. First is the steady, exponential amplification of magnetic energy. Second, once the magnetic energy has saturated on the smallest scale, the dynamo enters a slow linear or quadratic growth phase while the medium and large scale magnetic fields continue to be amplified \cite{choetal09, federrathetal11, schleicheretal13}. The third phase is the fully saturated phase once the magnetic field is saturated at all spatial scales.

The simulations are performed using the SPMHD code {\sc Phantom}, and the grid-based code {\sc Flash} \cite{fryxelletal00, dubeyetal08}, comparing results between the two distinct methods.

\subsection{Magnetised turbulence-in-a-box}

The simulations use a periodic box of length $L=1$. The initial density is $\rho=1$, velocity ${\bf v}=0$ and an isothermal equation of state is used, $P=c_{\rm s}^2 \rho$, with $c_{\rm s}=1$. The initial magnetic field is $B_z = 10^{-5}$ such that the plasma beta is $\beta = 10^{10}$. These simple initial conditions ensure that both codes start from identical states. The turbulence is driven by stirring the gas at large scales ($1 < k < 3$ peaked at $k=2$) through an added solenoidal force generated through an Ornstein-Uhlenbeck procedure \cite{ep88,schmidtetal09, federrathetal10}. This keeps the rms velocity of the gas steady at Mach 10 ($\mathcal{M}=10$), and maintains the reservoir of kinetic energy for the dynamo to draw upon. The simulation is run for 100 turbulent crossing times, defined as $t_{\rm c} = L / 2 \mathcal{M} c_{\rm s}$, capturing the full dynamo amplification phase and the statistical steady-state of the saturated phase. The simulations are performed for resolutions of $128^3$, $256^3$, and $512^3$ particles and grid cells.

\subsection{Results}

\begin{figure}
\centering
 \includegraphics[width=0.6\linewidth]{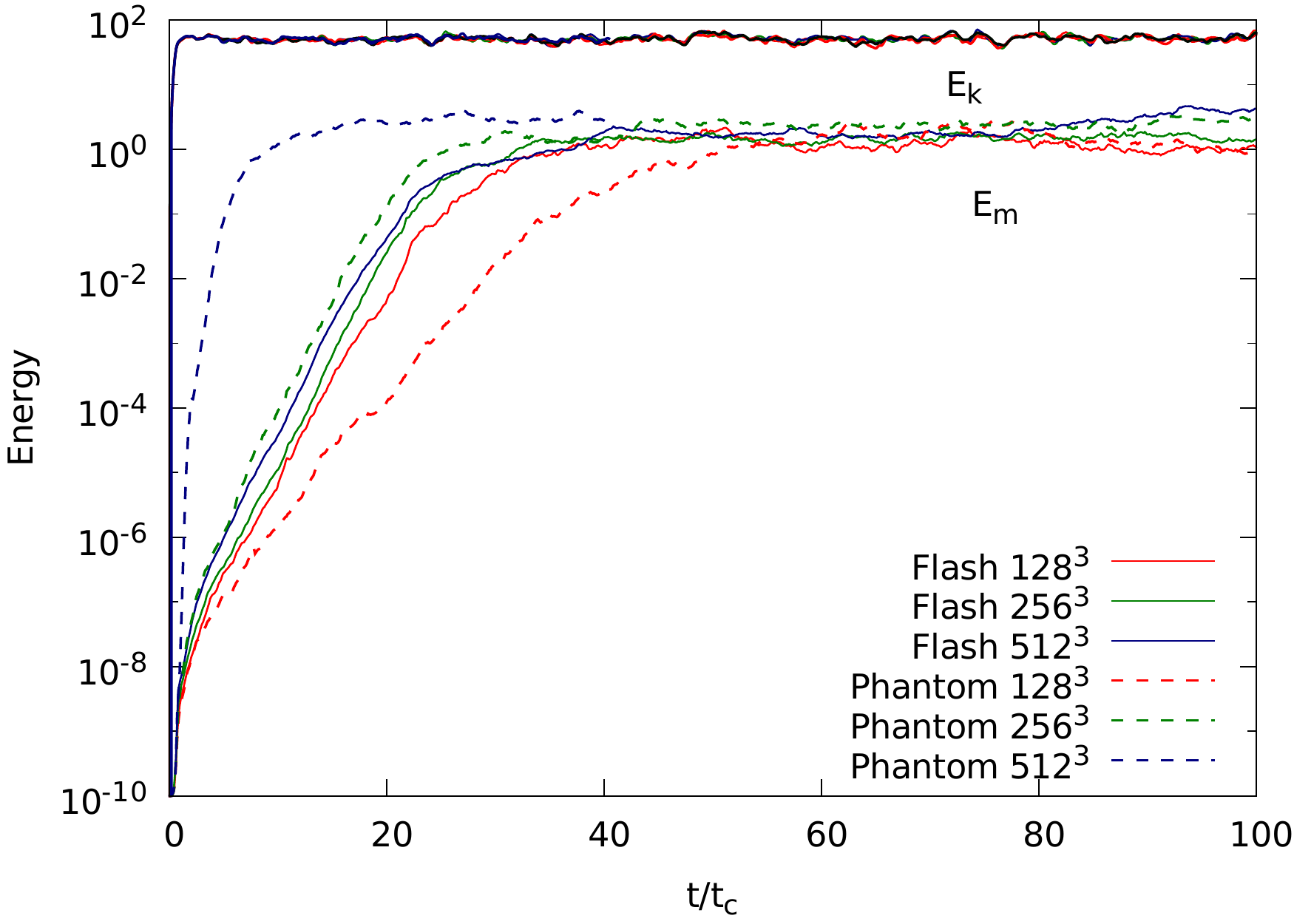}
\caption{The kinetic and magnetic energy for grid ({\sc Flash}, solid lines) and SPMHD ({\sc Phantom}, dashed lines) at resolutions of $128^3$ (red), $256^3$ (green), and $512^3$ (blue) grid points and particles. The kinetic energy for all calculations is maintained at a near constant level by the driving routine. Both codes amplify the magnetic energy at an exponential rate at early times, with a slow growth of magnetic energy as it nears saturation. For all calculations, the magnetic energy saturates at a similar level. {\sc Flash} has similar magnetic energy growth rates across the resolutions simulated, while {\sc Phantom} exhibits faster growth rates with increasing resolution. This resolution dependence is a consequence of the artificial dissipation terms used for shock capturing.}
\label{fig:en_mag}
\end{figure}

The evolution of the kinetic and magnetic energy for all calculations at all resolutions is shown in Fig.~\ref{fig:en_mag}. In all cases, the kinetic energy saturates within one crossing time ($E_{\rm k} \sim 50$), and is maintained it at a near constant level by the driving routine. The three phases of dynamo amplification of magnetic energy are evident: the steady, exponential amplification of magnetic energy, transitioning to the slow growth of magnetic energy ($E_{\rm m} \gtrsim 10^{-1}$), followed by the fully saturated magnetic field with a constant energy level ($E_{\rm m} \sim 10^0$). The exponential growth rate has little variation for the {\sc Flash} calculations, yet exhibits an increasingly higher growth rate for higher resolution with {\sc Phantom}. This can be traced to the differing shock capturing techniques used between the methods, as the numerical dissipation of the shock capturing is the primary factor in setting the growth rate of the dynamo for these simulations. All simulations saturate the magnetic energy at a similar level ($2$--$4$\% of the mean kinetic energy).

\begin{figure*}
\centering
\setlength{\tabcolsep}{0.0025\columnwidth}
\renewcommand{\arraystretch}{1.0}
\begin{tabular}{ccccl}
 \multicolumn{4}{c}{\sc Flash} & \\
  \includegraphics[height=0.2\textwidth]{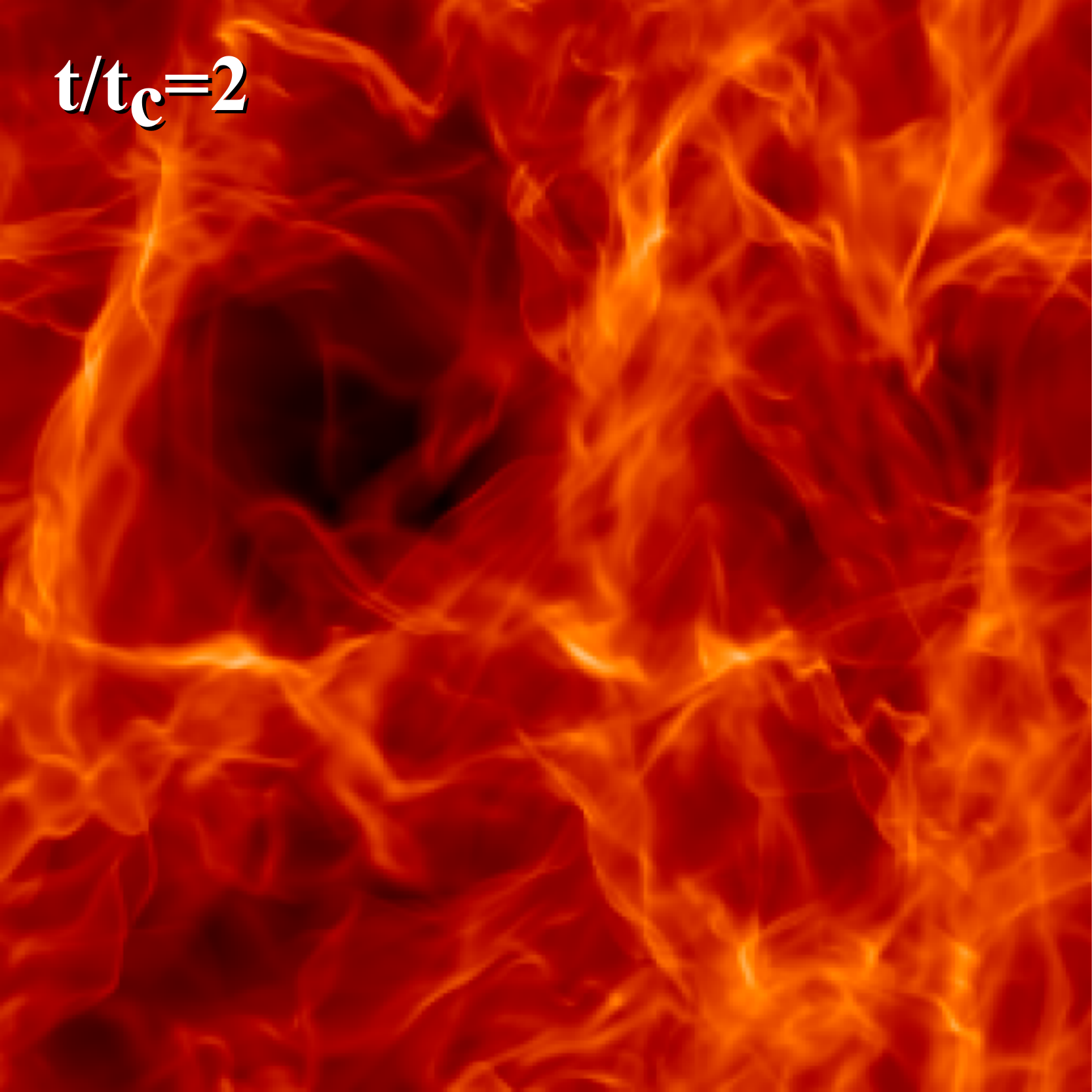}
& \includegraphics[height=0.2\textwidth]{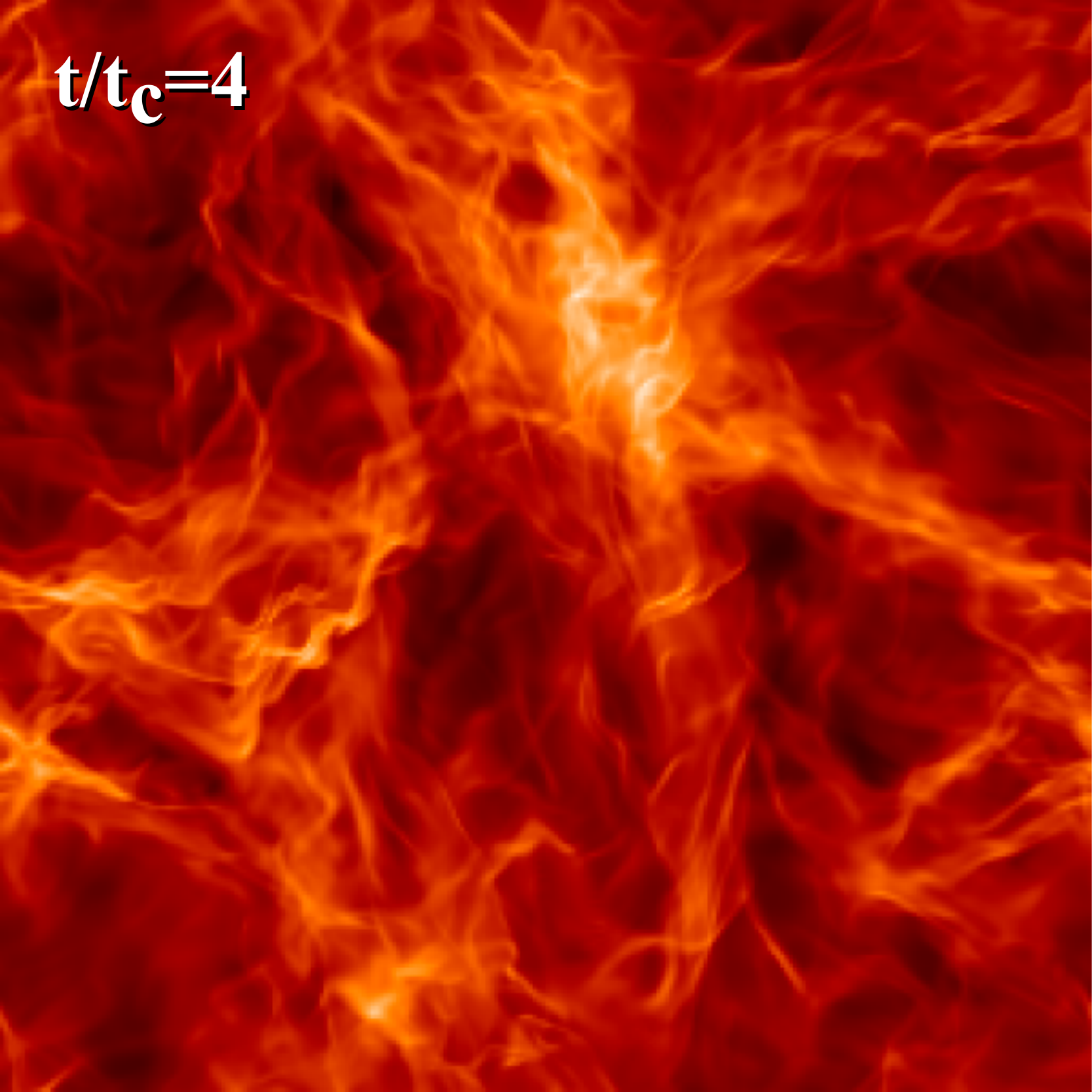}
& \includegraphics[height=0.2\textwidth]{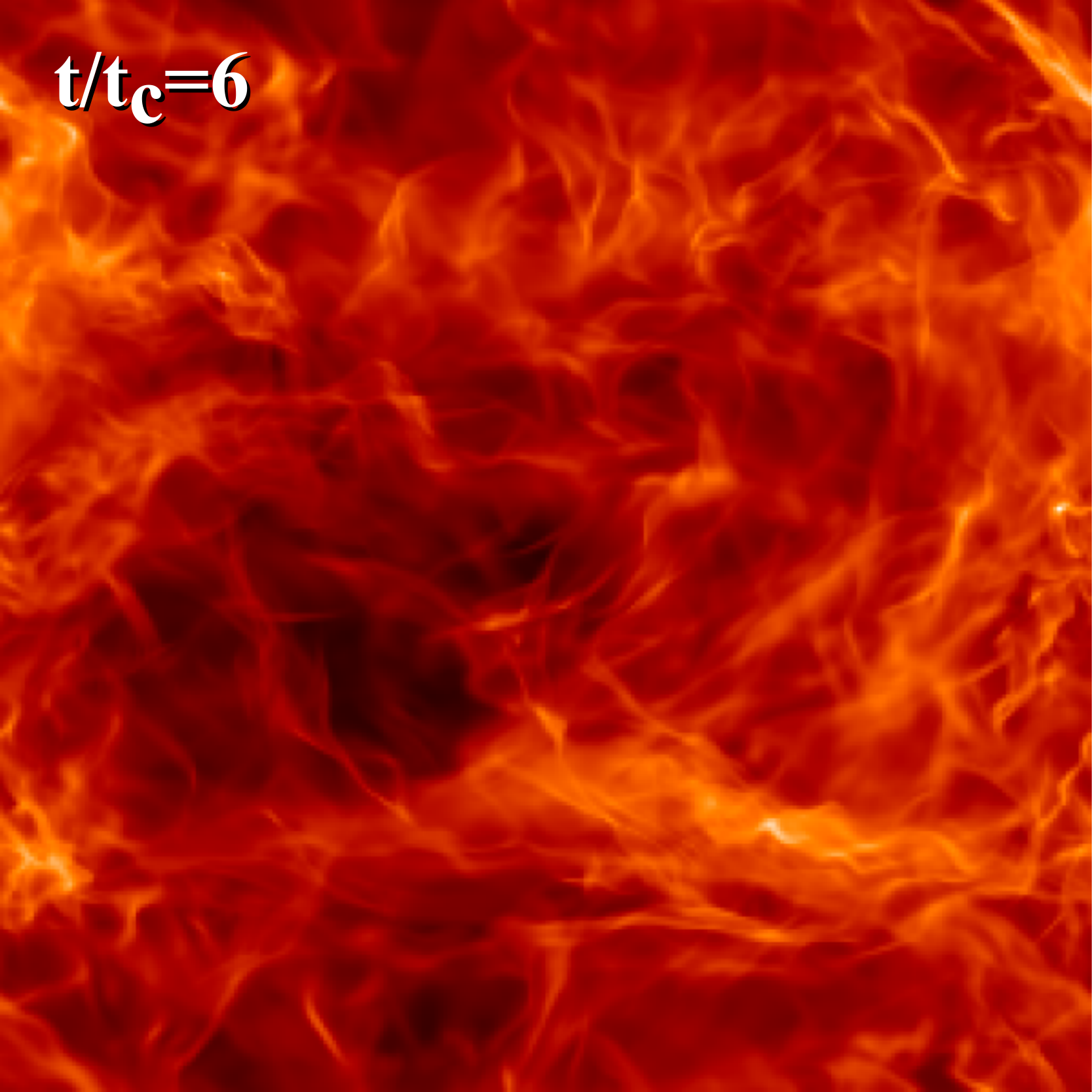}
& \includegraphics[height=0.2\textwidth]{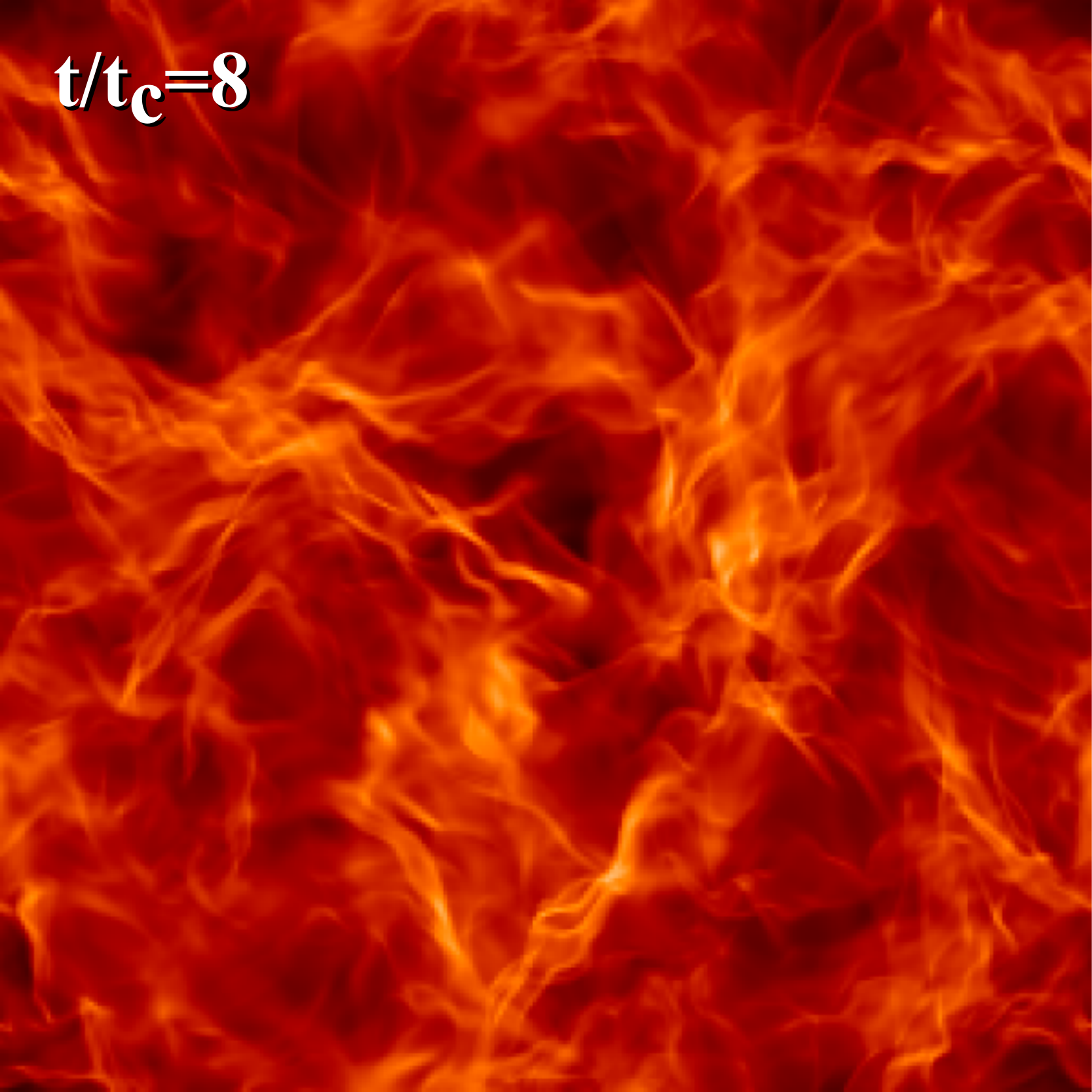} 
& \includegraphics[height=0.2\textwidth]{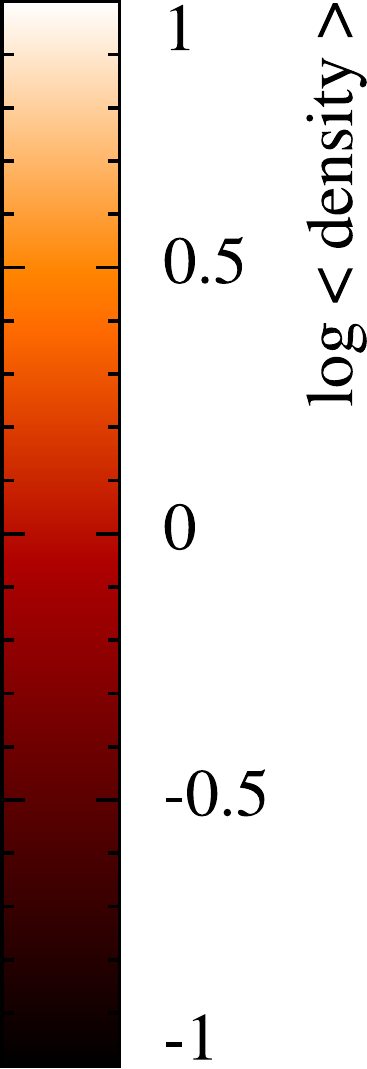} \\
  \includegraphics[height=0.2\textwidth]{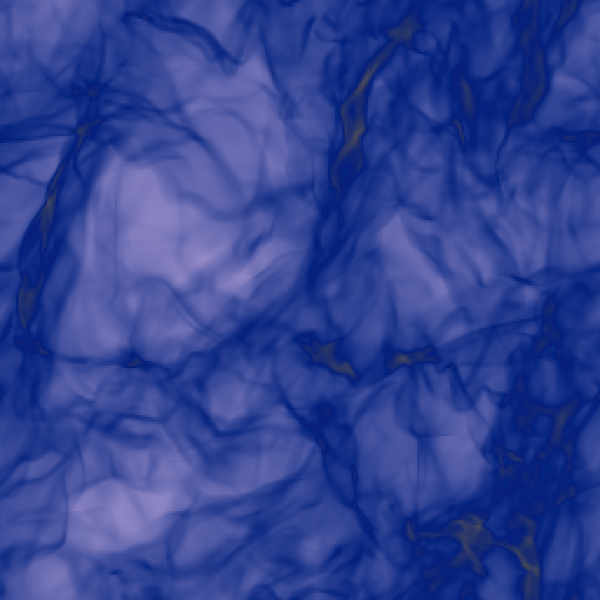}
& \includegraphics[height=0.2\textwidth]{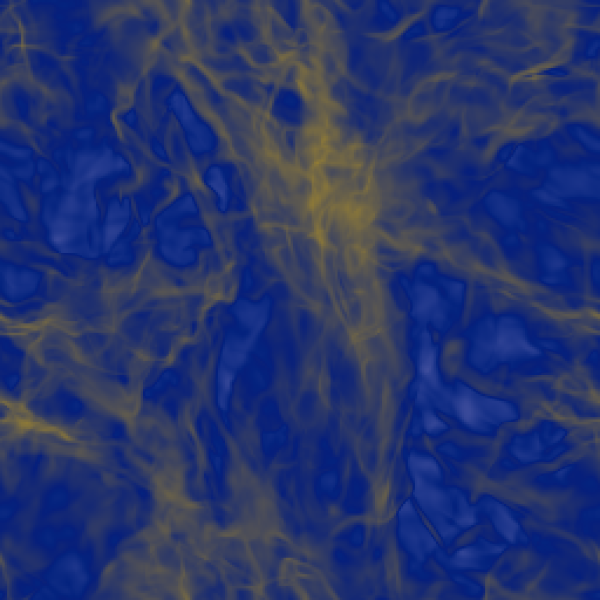}
& \includegraphics[height=0.2\textwidth]{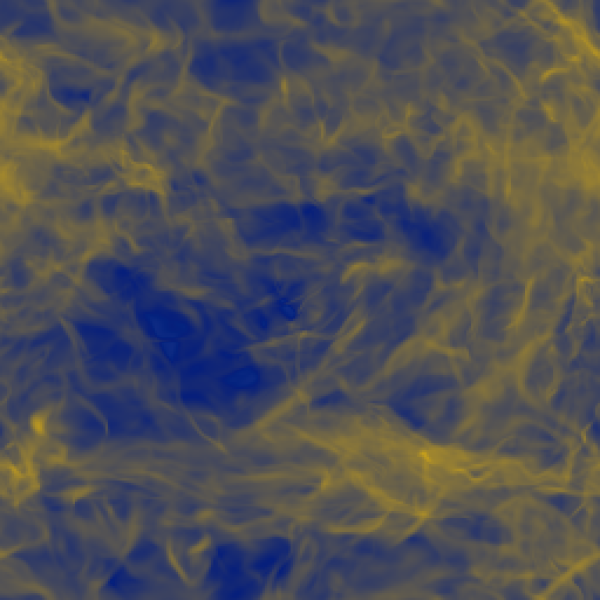}
& \includegraphics[height=0.2\textwidth]{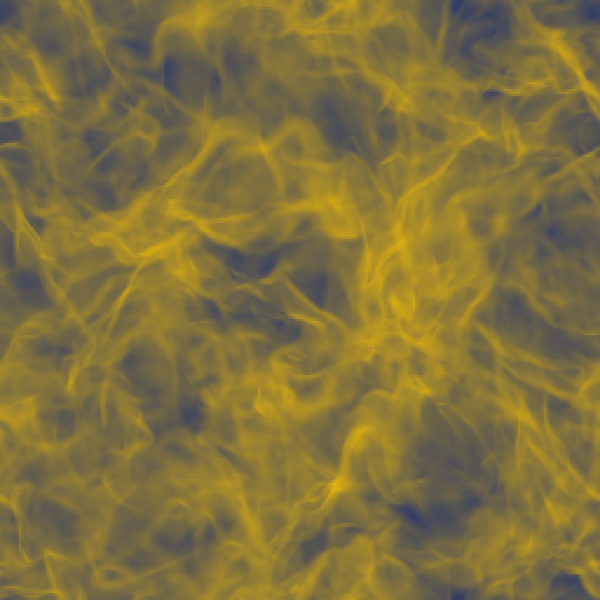}
& \includegraphics[height=0.2\textwidth]{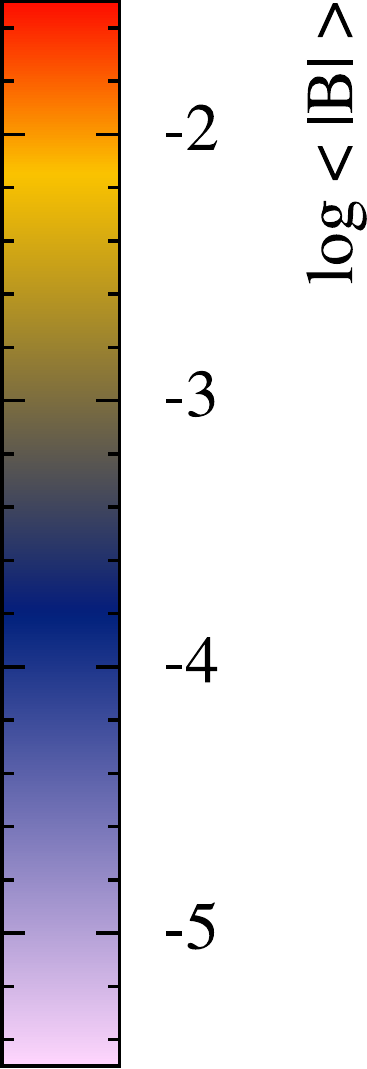}
\end{tabular} \\
\vspace{0.5cm}
\begin{tabular}{ccccl}
 \multicolumn{4}{c}{\sc Phantom} & \\
  \includegraphics[height=0.2\textwidth]{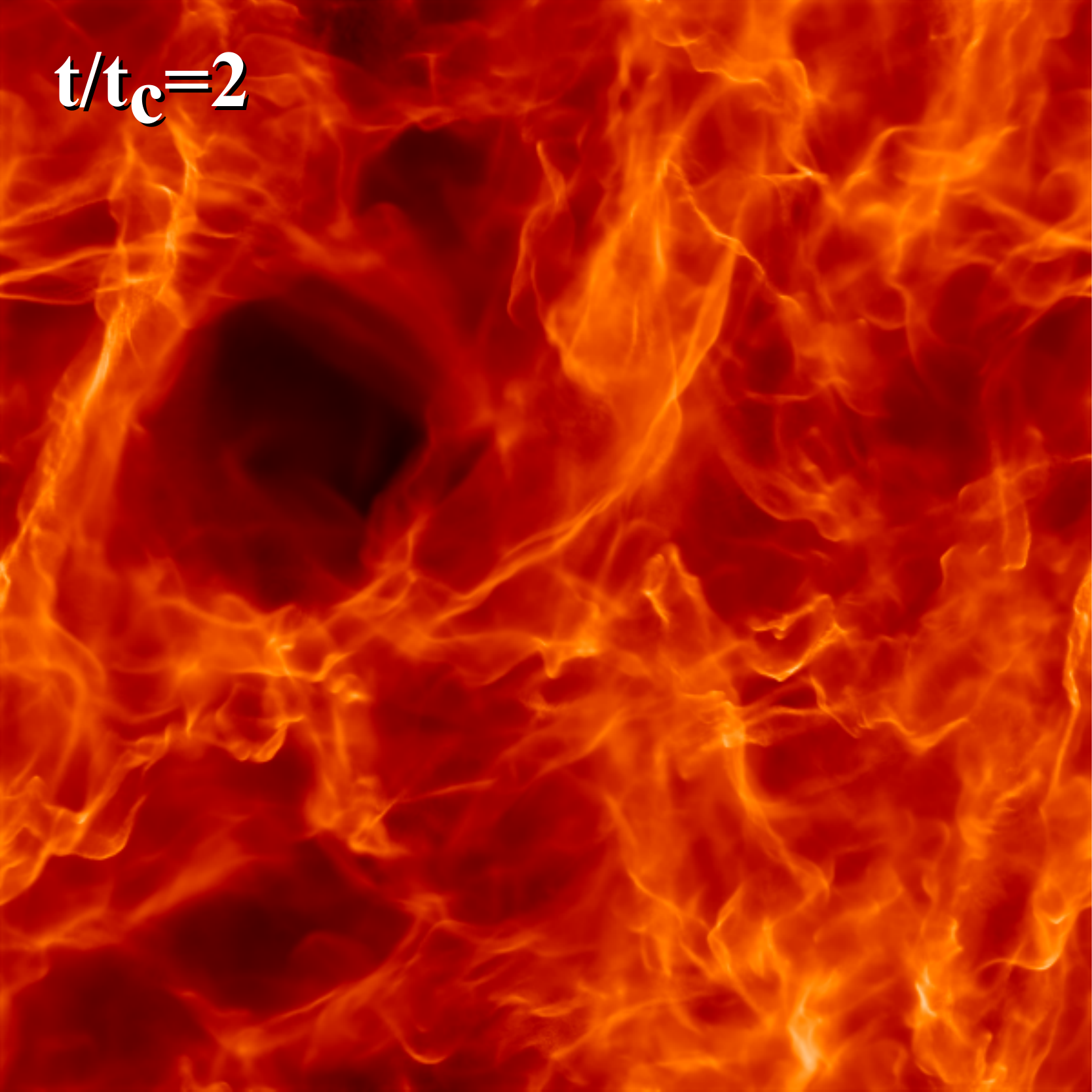}
& \includegraphics[height=0.2\textwidth]{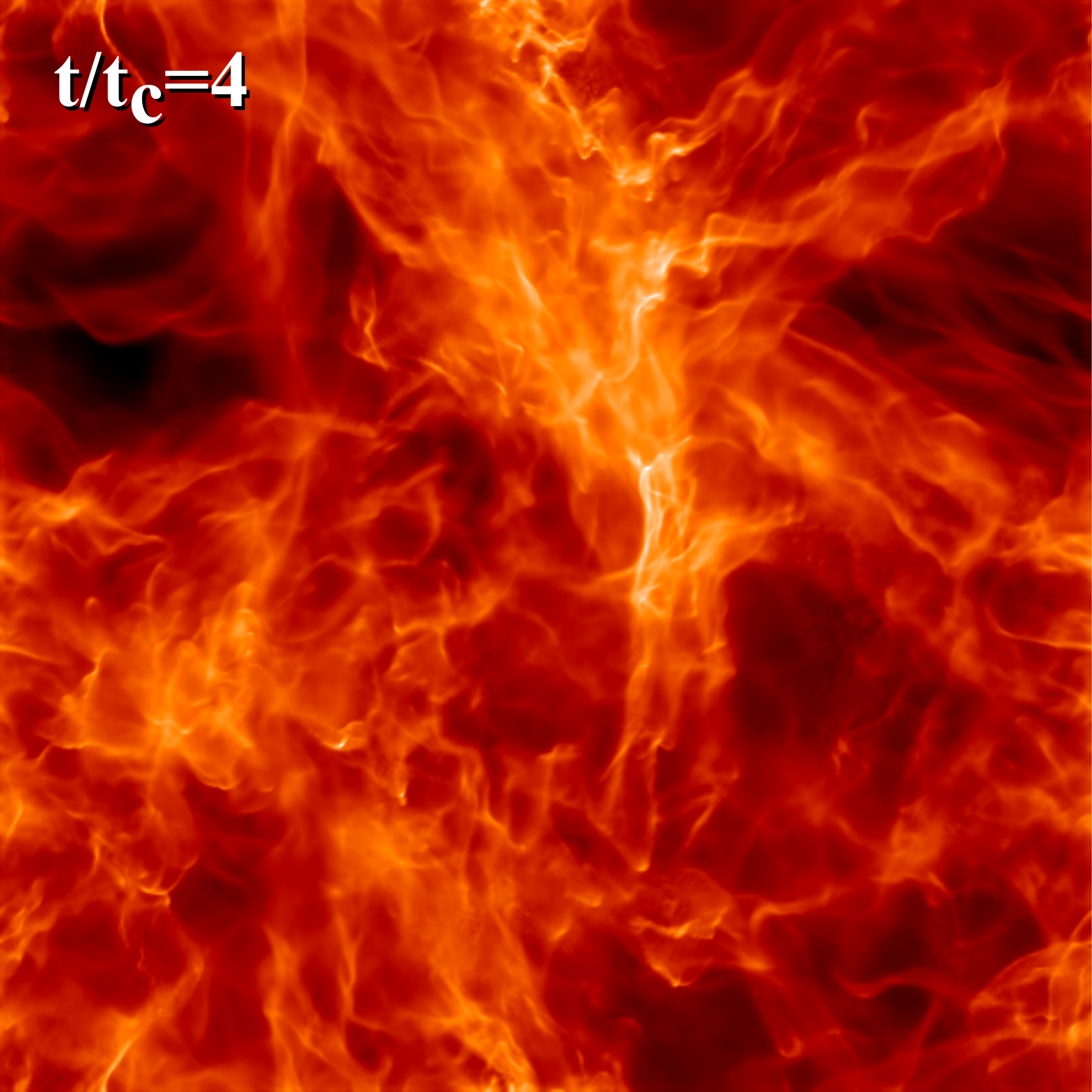}
& \includegraphics[height=0.2\textwidth]{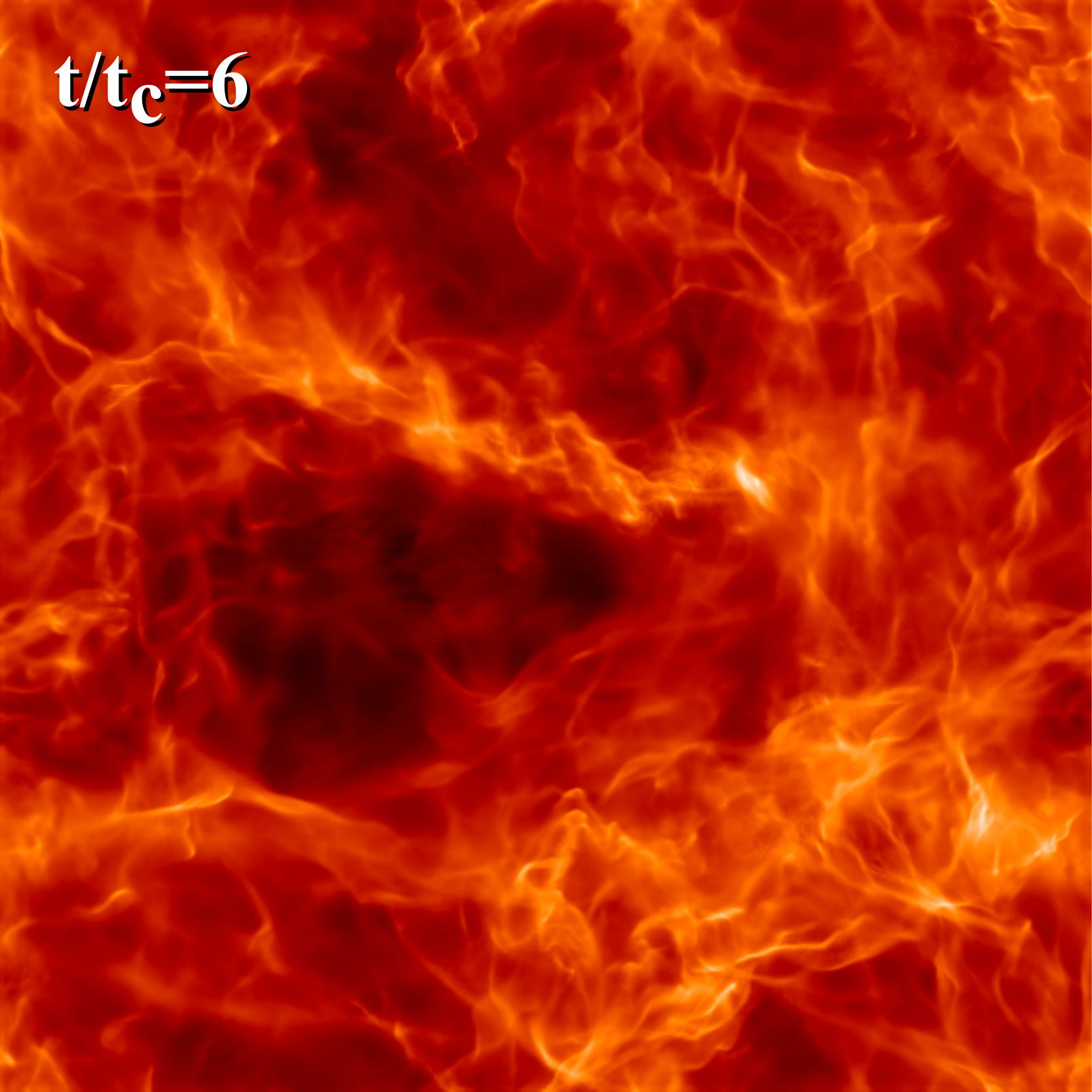}
& \includegraphics[height=0.2\textwidth]{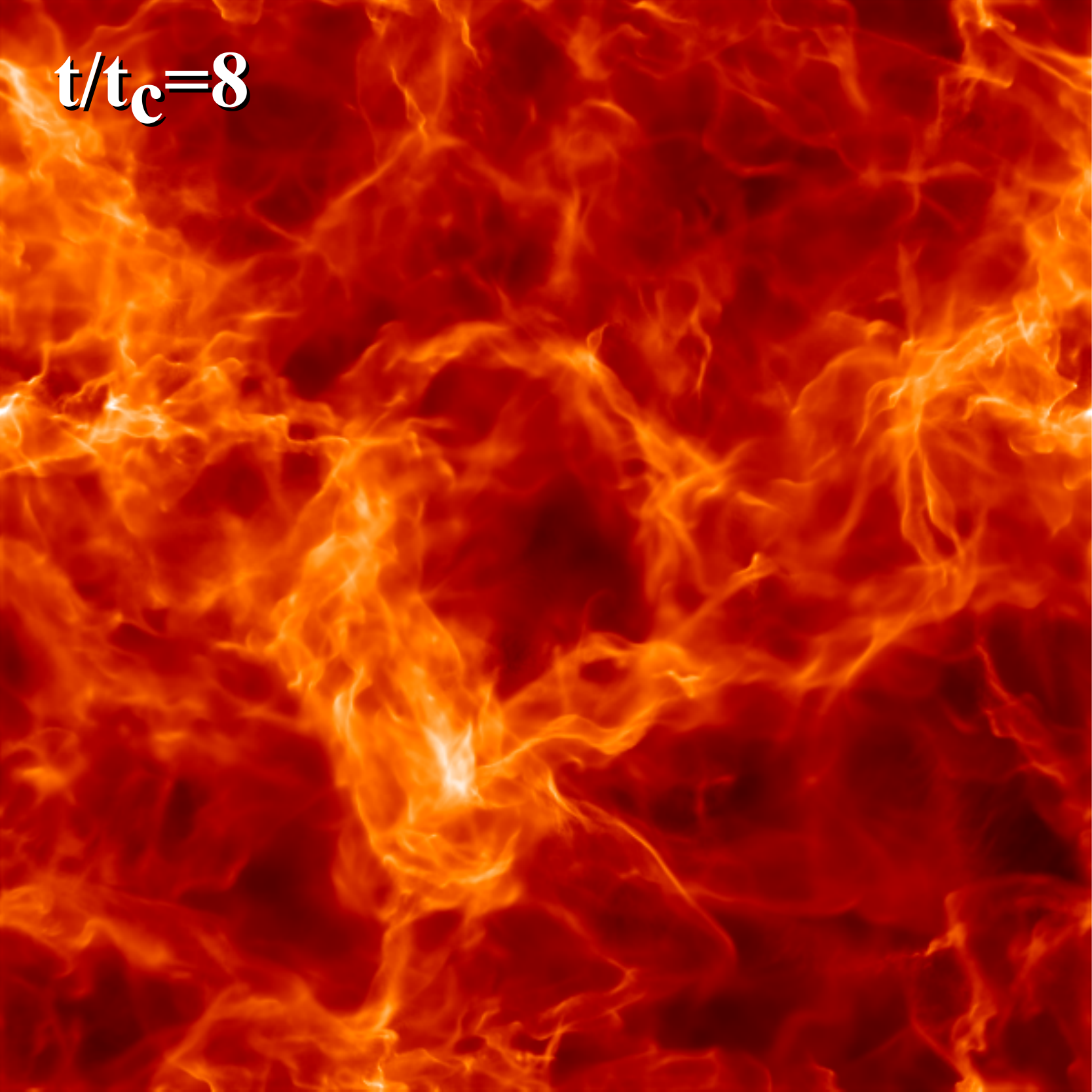}
& \includegraphics[height=0.2\textwidth]{cobar-column-rho-red.pdf} \\
  \includegraphics[height=0.2\textwidth]{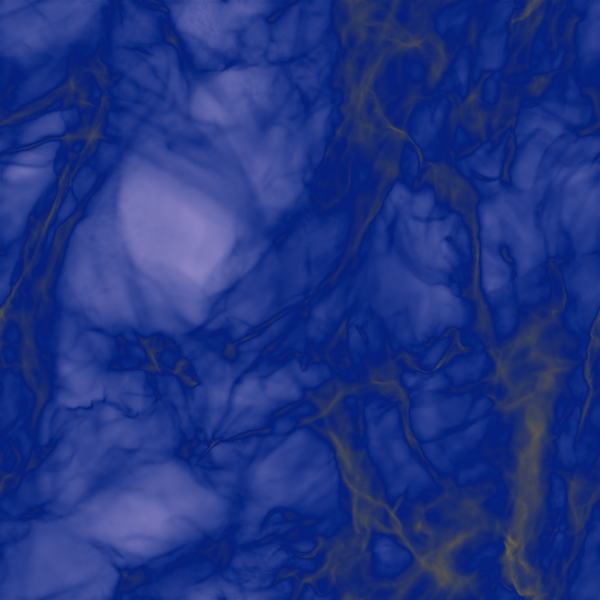}
& \includegraphics[height=0.2\textwidth]{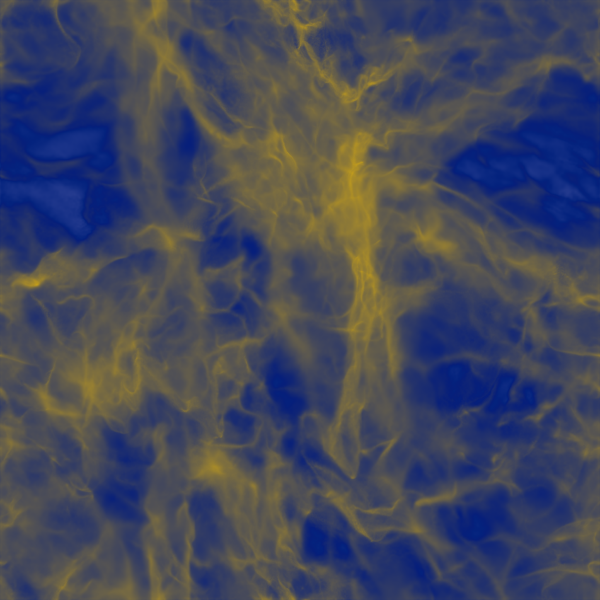}
& \includegraphics[height=0.2\textwidth]{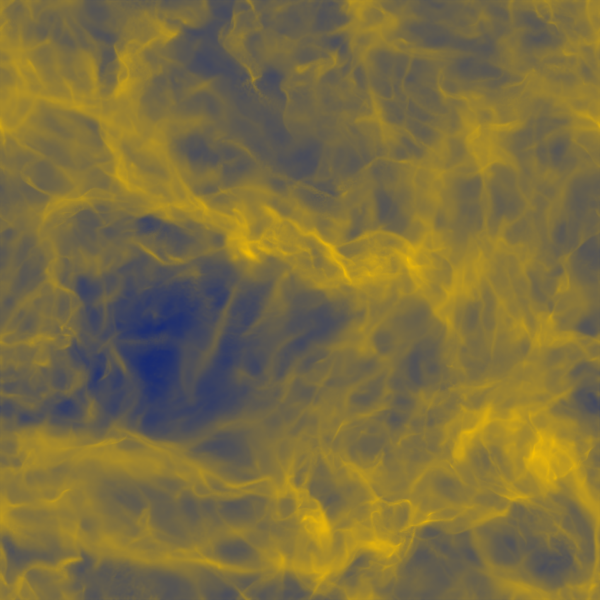} 
& \includegraphics[height=0.2\textwidth]{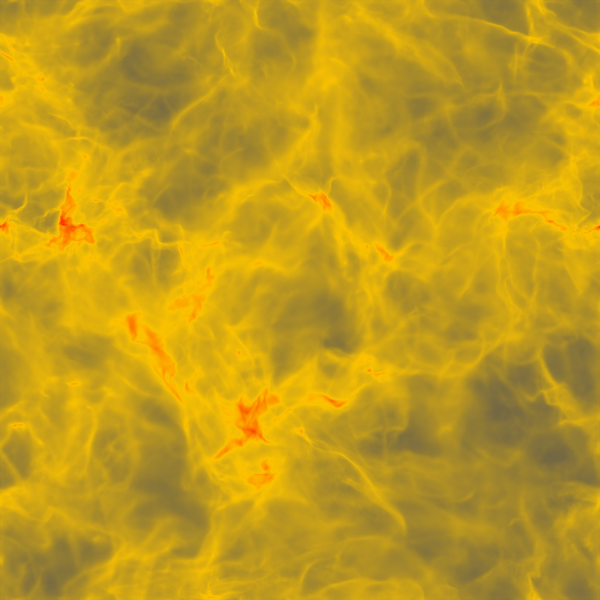} 
& \includegraphics[height=0.2\textwidth]{cobar-column-B-wbyr.pdf} 
\end{tabular}
\caption{$z$-column integrated $\rho$ and $\vert B \vert$, defined $<B> = \int \vert B \vert {\rm d}z / \int {\rm d}z$, for {\sc Flash} (top) and {\sc Phantom} (bottom) at resolutions of $256^3$ for $t/t_{\rm c}=2,4,6,8$. The density field has similar structure in both codes at early times, but diverge at late times due to the non-linear behaviour of the turbulence. The magnetic field is strongest in the densest regions, while the mean magnetic field strength throughout the domain increases with time.}
\label{fig:column-integrated}
\end{figure*}

Fig.~\ref{fig:column-integrated} shows the column density and column integrated magnetic field strength for the $256^3$ resolution calculations. At early times, both codes have similar density fields, but diverge at later times due to the non-linearity of the turbulence. The densest regions contain the strongest magnetic fields, yet the magnetic field throughout the whole domain is growing in strength.

\begin{figure}
 \centering
\includegraphics[width=0.8\columnwidth]{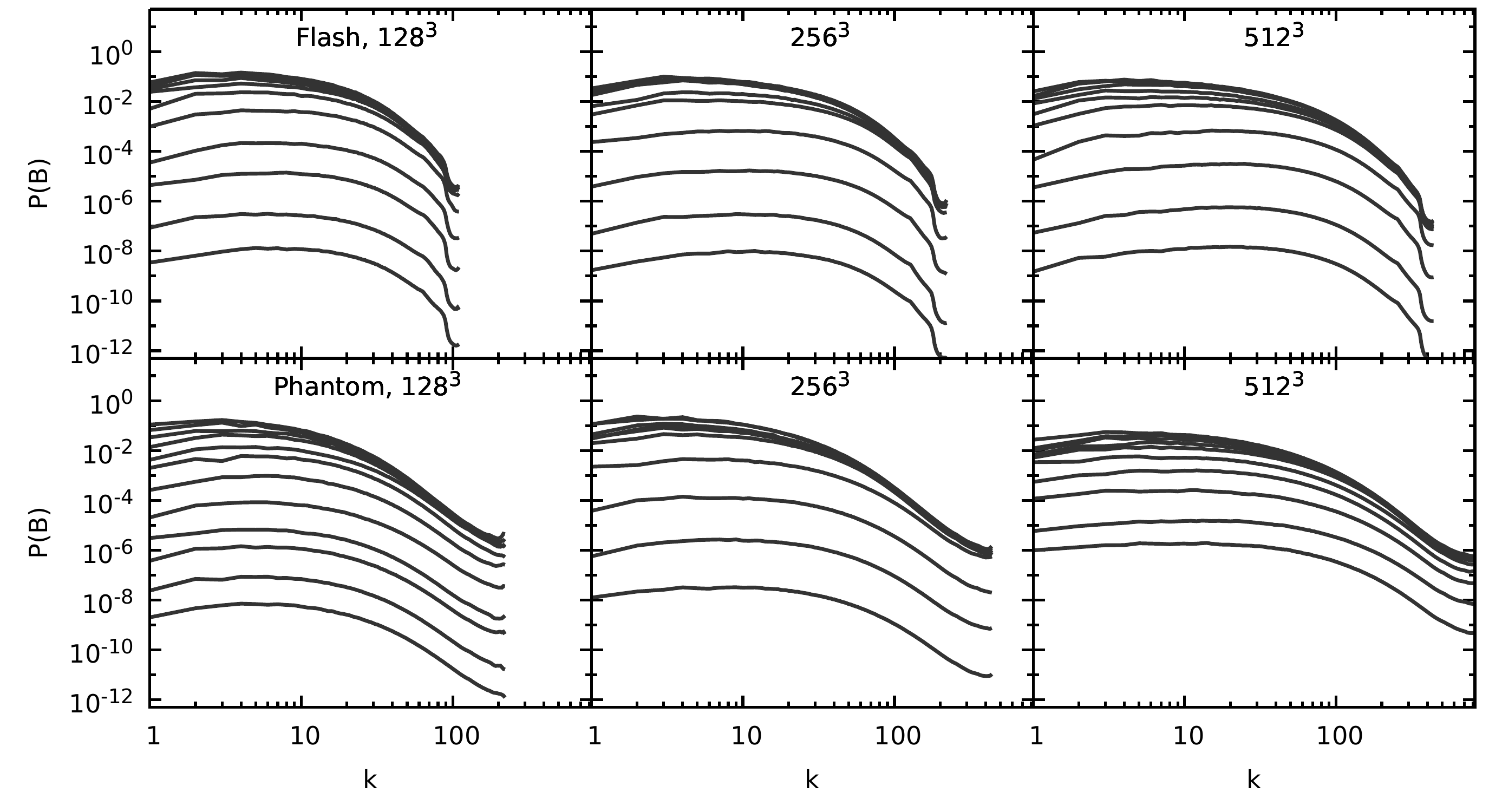}
\caption{Spectra of the magnetic energy during the growth phase for {\sc Flash} (top row) and {\sc Phantom} (bottom row) for resolutions of $128^3$, $256^3$, and $512^3$ (left to right). Each spectral line is sampled at intervals of $5t/t_{\rm c}$ up to $t/t_{\rm c}=50$, except for the $512^3$ {\sc Phantom} run which is sampled every $t/t_{\rm c}$ (from $t/t_{\rm c}=2$--$12$). Both codes exhibit the same qualitative behaviour: the magnetic energy spectra grow uniformly on all spatial scales with little change in shape, saturating first on the smallest scales with a time delay before the larger scales saturate.}
\label{fig:growthspectra}
\end{figure}

The growth of the magnetic field at different spatial scales can be quantified by looking at the power spectra of magnetic energy. As shown in Fig.~\ref{fig:growthspectra}, both codes exhibit the same behaviour for all resolutions, with the magnetic energy initially increasing uniformly at all spatial scales with little change in shape. This corresponds to the exponential growth phase, and is characteristic for the small-scale dynamo \cite{bs05, federrathetal14}. The power spectra converge first on the smallest spatial scales, at which point the dynamo enters the slow growth phase \cite{federrathetal14}. This may be seen by the time delay for the larger scales of the power spectra to converge.

 \begin{figure*}
\centering
\includegraphics[width=0.99\textwidth]{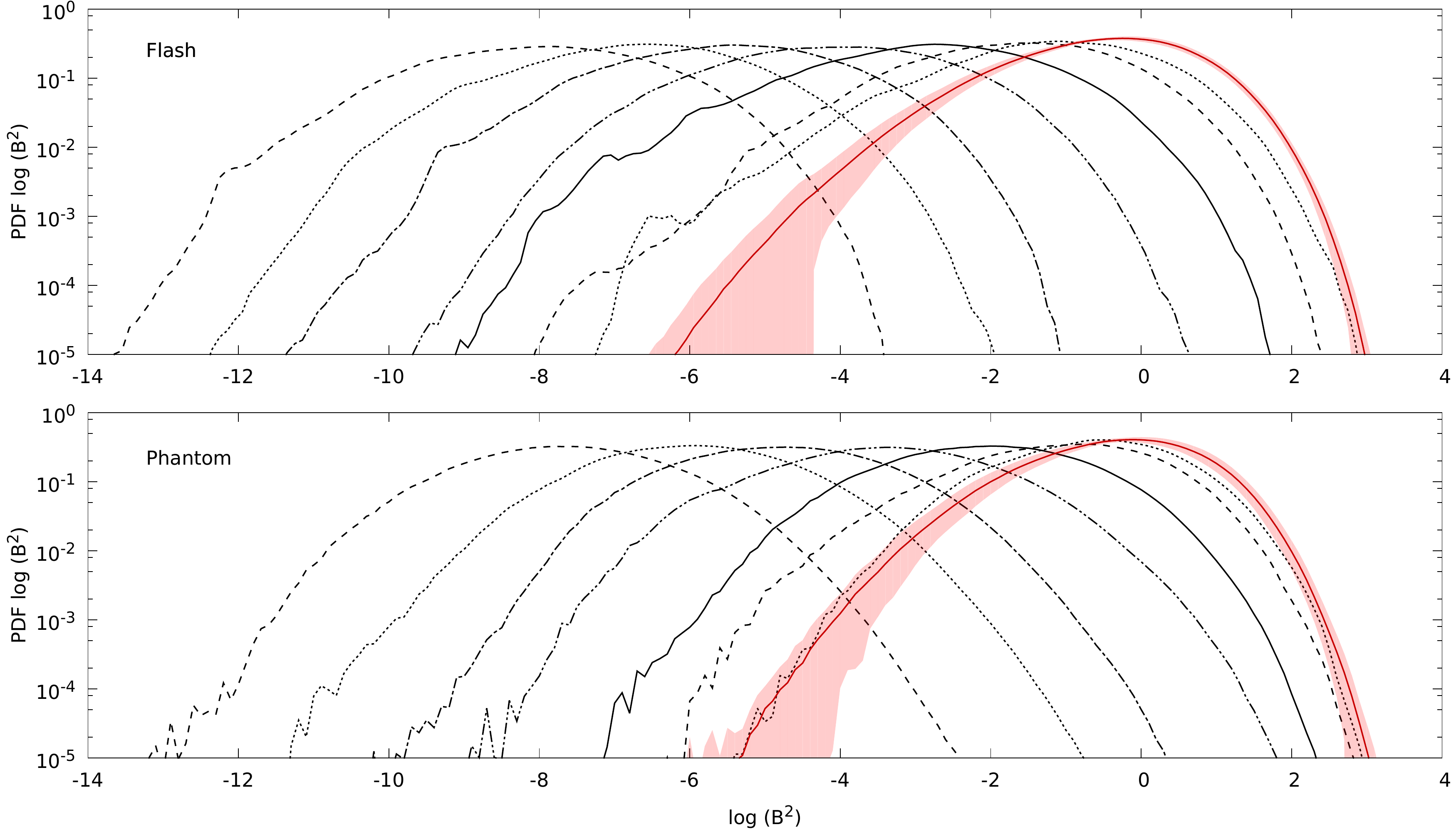}
\caption{PDF of $\log(B^2)$ during the growth phase, with the red line time averaged during the saturation phase (top panel: {\sc Flash}, bottom panel: {\sc Phantom}). The PDFs have log-normal distributions during the growth phase, maintaining their width while their peaks smoothly translate to higher magnetic field strengths. As the strongest magnetic fields saturate, the PDFs become lop-sided.}
\label{fig:bsqpdf}
\end{figure*}

Fig.~\ref{fig:bsqpdf} shows the probability distribution function (PDF) of $B^2$, with each curve representing a single snapshot in time and the red line the time-averaged PDF in the saturation phase. At early times, during the exponential growth phase, both {\sc Phantom} and {\sc Flash} have log-normal distributions of similar widths. The distributions maintain their width and shape as the magnetic field is amplified, with the mean increasing at a steady rate. As the dynamo nears saturation, the strongest magnetic fields no longer are amplified and the high-end tail of the distribution anchors in place. However, the low-end tail continues to be amplified (the slow growth phase), and the distribution takes on a lop-sided shape. Both codes exhibit this behaviour, consistent with previous numerical studies \cite{schekochihinetal04b, schekochihinetal04c}, and have similar maximum magnetic field strengths.


\section{Summary}
\label{sec:summary}

We have performed simulations of the small-scale dynamo amplification of magnetic fields in supersonic turbulence, comparing results between SPMHD and grid-based methods. The latest developments in SPMHD are utilised, namely the `constrained' hyperbolic divergence cleaning method to handle the divergence-free constraint of the magnetic field and a new switch to reduce numerical dissipation from artificial resistivity. The simulations are performed for 100 turbulent crossing times, allowing study of the full growth and saturation of magnetic energy due to dynamo amplification. Excellent agreement is found on the qualitative behaviour and properties of the magnetic field between SPMHD and the grid code. The exponential growth rate of magnetic energy has different resolution dependence, with SPMHD yielding faster growth rates for higher resolution, caused by the different approaches used to treat shocks. Overall, we conclude that SPMHD can correctly model the small-scale dynamo amplification of magnetic fields in supersonic turbulence.

\ack 
TST is supported by a CITA Post-doctoral Research Fellowship. The majority of this work was carried out while TST was still a PhD student, during which he was supported by an Australian Postgraduate Award and Endeavour International Postgraduate Research Scholarship. DJP is supported by ARC Future Fellowship FTI130100034, and acknowledges funding provided by the ARCÕs Discovery Projects (grant no. DP130102078). CF acknowledges funding provided by the Australian Research CouncilÕs Discovery Projects (grants DP130102078 and DP150104329). This research was undertaken with the assistance of resources provided at the Multi-modal Australian ScienceS Imaging and Visualisation Environment (MASSIVE) through the National Computational Merit Allocation Scheme supported by the Australian Government. We gratefully acknowledge the J{\"u}lich Supercomputing Centre (grant hhd20), the Leibniz Rechenzentrum and the Gauss Centre for Supercomputing (grant pr32lo), the Partnership for Advanced Computing in Europe (PRACE grant pr89mu), and the Australian National Computing Infrastructure (grant ek9). Some figures were created using the interactive SPH visualisation tool {\sc Splash} \cite{splash}.

\bibliographystyle{iopart-num}
\bibliography{astronumbib}

\end{document}